\documentclass[preprint,showpacs,preprintnumbers,amsmath,amssymb,nofootinbib,floatfix]{revtex4}

\usepackage{graphicx}
\usepackage{dcolumn}
\usepackage{bm}

\begin{document}
\baselineskip=13pt

\preprint{}

\title{Giant halo in relativistic and non-relativistic approaches}

\author{ J.~Terasaki$^1$, S.~Q.~Zhang$^{1,2}$, S.~G.~Zhou$^{2,3}$ and J.~Meng$^{1-3}$}
\affiliation{$^{1}$ School of Physics, Peking University, Beijing 100871,
China \\
$^{2}$ Institute of Theoretical Physics, Chinese Academy of Science,
Beijing 100080, China \\
$^{3}$ Center of Theoretical Nuclear Physics, National Laboratory of
Heavy Ion Accelerator, Lanzhou 730000, China}


\date{\today}

\begin{abstract}
\baselineskip=13pt
The phenomena of giant halo and halo of neutron-rich even-Ca
isotopes are investigated and compared in the framework of the
relativistic continuum Hartree-Bogoliubov (RCHB) and
non-relativistic Skyrme Hartree-Fock-Bogoliubov (HFB) calculations.
With two parameter sets for each of the RCHB and the Skyrme HFB calculations,
it is found
that although halo phenomena exist for Ca isotopes near
neutron drip line in both calculations,  the halo of
the Skyrme HFB calculations starts at a more neutron-rich nucleus
than that of the RCHB calculations, and the RCHB calculations have larger
neutron root-mean-square (rms) radii systematically in
$N \geq 40$ than those of the Skyrme HFB calculations.
The former difference comes from difference in shell structure.
The reasons for the latter can be partly explained by the
neutron 3$s_{1/2}$ orbit, which causes more than 50 \% of the difference
 among the four calculations for neutron rms radii at $^{66}$Ca.
\end{abstract}


\maketitle


\section{\label{sec:introduction} Introduction}
The study of exotic nuclei with extreme $N/Z$ ratios has attracted
worldwide attention  since the first discovery  of halo in
$^{11}$Li \cite{Tan85}. Usually for exotic nuclei, their Fermi
surfaces are very close to the continuum threshold,  and the valence
nucleons could be easily scattered to the continuum states due to
the pairing correlations \cite{Men96,Men98c}. Thus, theories which
can properly handle the pairing and continuum states are needed to
describe the properties of exotic nuclei and to understand the
interference of the continuum-energy states to bound many-body
systems and effects of large neutron excess.

It is well known that the BCS approximation for the pairing
correlations suitable for stable nuclei cannot give correct wave
functions near the drip lines, see e.g.~\cite{Dob96}. Taking into
account the pairing correlations by the Bogoliubov transformation,
relativistic and/or non-relativistic mean field approaches have been
extensively used to describe and predict halo phenomena in exotic
nuclei.

The status of the studies of the halo using the relativistic
continuum Hartree-Bogoliubov (RCHB) or the non-relativistic
Hartree-Fock-Bogoliubov (HFB) method include:  the predictions of
halo in the Ne ~\cite{Pos97,Lal98}, Na ~\cite{Men98b,Lal98},  Ca
\cite{Men02} and Zr \cite{Men98} isotopes near the neutron drip
line. More details can be found in recent review papers, e.g.,
\cite{Men05}. As there are more than two particles in the weakly
bound or the positive-energy region in Ca and Zr isotopes near the
neutron drip line, the halo phenomena for Ca and Zr are addressed as
giant halo \cite{Men98,Men02}. The proton-magic nuclei from O and Pb
nuclei are also systematically surveyed, and it is found that the
halos signature in Zr is the clearest in Ref.~\cite{Zha03}. There is
no evidence for halo in Ni, Sn and Pb isotopic chains, similarly as
in Ref.~\cite{Rin02}.

In the non-relativistic HFB approaches, Skyrme calculations with the
parameter set SLy4 predicts the halo in Sn and Ni~\cite{Miz00}. It
is also claimed that the pairing gaps have an effect to reduce the
halo at the neutron drip line~\cite{Miz00,Ben00,Dob01}.

For comparison with the halos predicted in relativistic
approaches \cite{Men98,Men02,Zha03}, it is necessary to investigate
the Ca and Zr in non-relativistic approaches, which is very
important but still missing in the literature. Therefore the present
paper is dedicated to investigate the neutron-rich even Ca isotopes
up to the neutron drip line.  The results from both relativistic and
non-relativistic mean field approaches taking into account the
pairing correlations by the Bogoliubov transformation will be
presented and compared in detail.

The numerical details used in both calculations are explained in
Sec.~\ref{sec:calculation}, and then the neutron-number dependence
of the energy and the rms radii is examined in Sec.~\ref{sec:halo
structure}.  The properties of single particles are discussed in
Sec.~\ref{sec:66ca} in order to understand the halo structure. A
brief summary is given in Section \ref{sec:conclusion}.

\section{\label{sec:calculation} Method of calculation}

The detailed formulations of the RCHB method and the HFB method can
be found respectively in Refs.~\cite{Men96,Men98c}
and \cite{Dob84,Benn05}. Both calculations are based on the
coordinate representation and use a box size 20 fm with the
spherical symmetry assumed and a mesh size 0.1 fm. The quasiparticle
states are obtained up to 120 MeV (165--190 radial wave functions in
$^{66}$Ca), and all of these states are used for calculating
potentials. The maximum angular momentum of the quasiparticles
$j_{\rm max}$ is $\frac{13}{2}$ (RCHB) and $\frac{15}{2}$ (Skyrme
HFB). The parameter sets NL-SH \cite{Sha93} and PK1 \cite{Lon04} (
SkM$^\ast$ \cite{Bar82} and SLy4 \cite{Cha98} ) are used for the
RCHB ( Skyrme HFB ) calculations. For the pairing part, a
surface-type delta interaction \cite{Dob96,Men98} is used with the
density parameter $\rho_0 = $ 0.152 fm$^{-3}$ and 0.160 fm$^{-3}$
for the RCHB and the Skyrme HFB calculations, respectively.

The strength of the pairing interaction in mean-field calculations
is normally determined so as to reproduce the pairing gaps obtained
from odd-even mass differences (see e.g.~Ref.~\cite{Dob01b}). While
for nuclei close to the neutron drip line, the pairing energy
obtained using the Gogny interaction in the pairing channel is
adopted to fix the strength of surface-type delta pairing
interaction due to the missing of experimental data.
The $V_0 $ is $-$325 MeV$\,$fm$^3$ in the RCHB calculations, and
that of the Skyrme HFB calculations is
$-$365 ($-$330) MeV$\,$fm$^3$ in the SLy4 (SkM$^\ast$) calculations.
The average pairing evergy of the PK1 and NL-SH calculaitons $-$8.2 MeV
at $^{66}$Ca is used for determining the $V_0$ of the Skyrme-HFB calculaitons.
We checked that when $j_{\rm max}=\frac{13}{2}$ was
used in the Skyrme HFB calculations, the rms radii did not change
with the re-adjusted $V_0$ from the pairing energy. 

\begin{figure}
\includegraphics[width=0.45\textwidth]{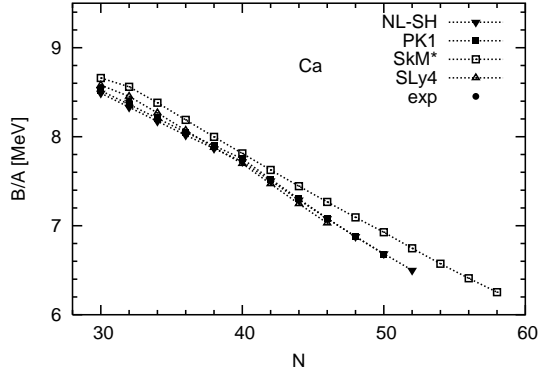}
\caption{\label{fig:e}
\baselineskip=13pt
Binding energy per nucleon $B/A$ of even Ca
in RCHB calculations with NLSH and PK1 and the Skyrme HFB
calculations with SkM$^\ast$ and SLy4 compared with the experimental
data \cite{Aud03} at $N=30$ and 32. For the details, see the text.
}
\end{figure}
\begin{figure}
\includegraphics[width=0.45\textwidth]{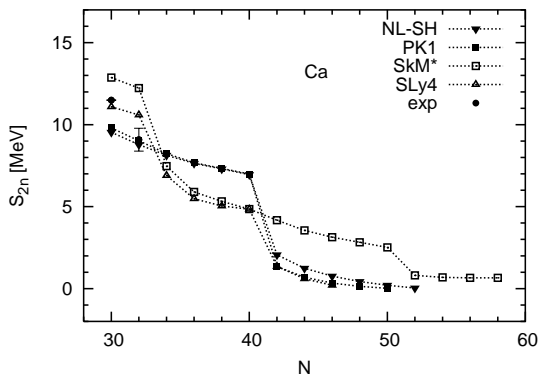}
\caption{\label{fig:s2n}
\baselineskip=13pt
Two-neutron separation energy $S_{2n}$ of
even Ca in RCHB calculations with NLSH and PK1 and the Skyrme HFB
calculations with SkM$^\ast$ and SLy4 compared with the experimental
data \cite{Aud03}.
Note that $S_{2n}$ of SLy4 are very close to those of PK1
in $42 \leq N \leq 48$.
}
\end{figure}

\section{\label{sec:halo structure} Neutron-number dependence of
the halo structure} Figures \ref{fig:e} and \ref{fig:s2n} show the
binding energy per nucleon $B/A$ and the two-neutron separation
energy $S_{2n}$ in RCHB calculations with NLSH and PK1 and the
Skyrme HFB calculations with SkM$^\ast$ and SLy4, respectively. The
most important differences between these calculations are location
of the two-neutron drip line and a shell gap at $N=40$. The RCHB
calculations show the two-neutron drip line at $N=50$ (PK1) or 52
(NL-SH), and the prediction of the Skyrme HFB calculations is
$N=46$ (SLy4) or 58 (SkM$^\ast$). The RCHB and the SLy4 calculations
have clear the shell gap at $N=40$ (Fig.~\ref{fig:s2n}), while the
SkM$^\ast$ calculation does not show at all. The SkM$^\ast$
calculation has a small extra downward behavior of $S_{2n}$ at
$N=50$, however, it is too small to call a shell gap. Therefore, no
new shell gap is predicted in $N>40$, in any of the calculations.
The two RCHB calculations give the results rather close to each
other both in $B/A$ and $S_{2n}$, and the two Skyrme HFB
calculations have a difference of 0.1--0.2 MeV in $B/A$. The
$S_{2n}$ of SLy4 is close to those of the SkM$^\ast$ calculation in
$34 \leq N \leq 40 $ and then coincides with that of PK1 in $42 \leq
N \leq 48$. The experimental data are available up to $N=32$
currently \cite{Aud03}. The RCHB calculations reproduce well the
data of $B/A$ at $N=30$ (8.550 MeV) and 32 (8.396$\pm$0.013 MeV) and
$S_{2n}$ at $N=32$ (9.081$\pm$0.699 MeV). For $S_{2n}$ at $N=30$,
the SLy4 calculation gives a value very close to the measured one
(11.499$\pm$0.008 MeV).

\begin{figure}
\includegraphics[width=0.45\textwidth]{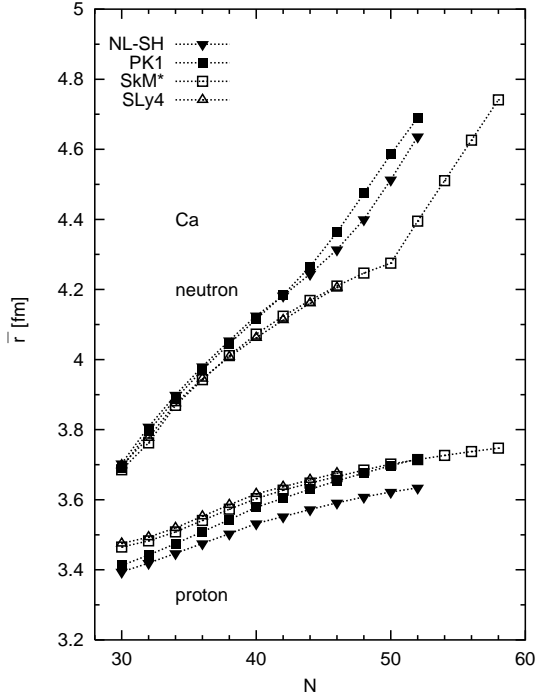}
\caption{\label{fig:radii3np}
\baselineskip=13pt
The neutron and proton $rms$ radii
$\bar{r}_n$ and $\bar{r}_p$ of even-Ca isotopes in RCHB calculations
with NLSH and PK1 and the Skyrme HFB calculations with SkM$^\ast$
and SLy4 in the neutron-rich region. }
\end{figure}

\begin{figure}
\includegraphics[width=0.45\textwidth]{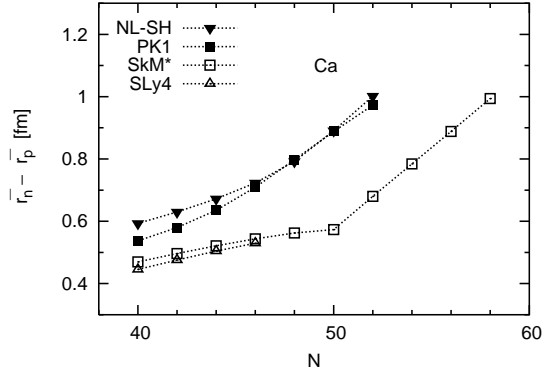}
\caption{\label{fig:radii2b}
\baselineskip=13pt
The difference between neutron and
proton $rms$ radii $\bar{r}_n-\bar{r}_p$ in RCHB calculations with
NLSH and PK1 and the Skyrme HFB calculations with SkM$^\ast$ and
SLy4 for even-Ca isotopes with $N \geq 40$. }
\end{figure}

\begin{figure}
\includegraphics[width=0.45\textwidth]{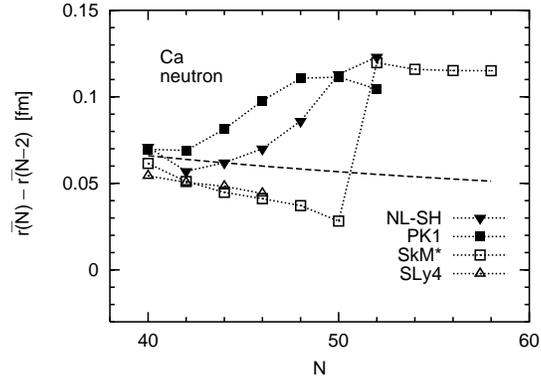}
\caption{\label{fig:del_radii_n}
\baselineskip=13pt
The difference of neutron $rms$
radii between neighboring even-Ca isotopes with $N \geq 40$
$\bar{r}_n(N)-\bar{r}_n(N-2)$ in RCHB calculations with NLSH and PK1
and the Skyrme HFB calculations with SkM$^\ast$ and SLy4 in the
neutron-rich region. The dashed line was obtained from
$\bar{r}_n(N)=1.139 N^{1/3}$ fm \cite{Men02}. }
\end{figure}

Figure \ref{fig:radii3np} shows systematics of the rms radii of
neutrons $\bar{r}_n$ and protons $\bar{r}_p$ of even-Ca from $N=30$
to the two-neutron drip lines. The increase in the curvature of
$\bar{r}_n$ indicates the halo. (For attempts to define the
criterion of halo or the halo size, see
Refs.~\cite{Miz00,Men98b,Che05} and references therein.) For Ca, the
number of nucleons in the positive-energy region $N_h$ of
$^{62-72}$Ca is 0.6--2.2, of which average is 1.7, in the NL-SH
calculation \cite{Men02}, and the corresponding value of the
SkM$^\ast$  calculation is always smaller than 0.5 (the average
0.27). As there are more than two nucleons in the weakly bound
orbits and the positive-energy region for these nuclei and also halo
phenomena in neighboring nuclei with incremental neutrons, the halo
in Ca isotopes are refereed as giant halo, as in
Refs.~\cite{Men98,Men02}. The SkM$^\ast$ calculation has the halo
from $N=52$, and the RCHB calculations show the gradual occurrence
of the giant halo. Apparently, the starting nucleus of the halo of
SkM$^\ast$ corresponds to the extra lowering of the $S_{2n}$. The
SLy4 calculation does not have the halo, because the particle-stable
region ends at $N=46$. The large difference between $\bar{r}_n$ and
$\bar{r}_p$ shown in Fig.~\ref{fig:radii3np} is identified with
neutron skin in the region with no halo. $\bar{r}_n-\bar{r}_p$ is
displayed in Fig.~\ref{fig:radii2b}, in which the differences
between the parameter sets are clear; the curve of SLy4 is parallel
to that of SkM$^\ast$, and PK1 has a larger curvature than the other
parameter sets in $N \leq 50$. An important aspect of the halo is
how $\bar{r}_n$ changes when two more neutrons are added, thus it is
worth showing the difference $\bar{r}_n(N)-\bar{r}_n(N-2)$
(Fig.~\ref{fig:del_radii_n}). When there is no halo, two additional
neutrons increase $\bar{r}_n$ by 0.05 fm, and the increase rises to
0.11 fm in the halo nuclei. In all of
Figs.~\ref{fig:radii3np}--\ref{fig:del_radii_n}, the difference in
$\bar{r}_n$ is clear between the RCHB and the Skyrme HFB
calculations. It is to be noted that $\bar{r}_p$ is not constant,
reflecting the self-consistency of the neutrons and protons
interaction.

\begin{figure}
\includegraphics[width=0.45\textwidth]{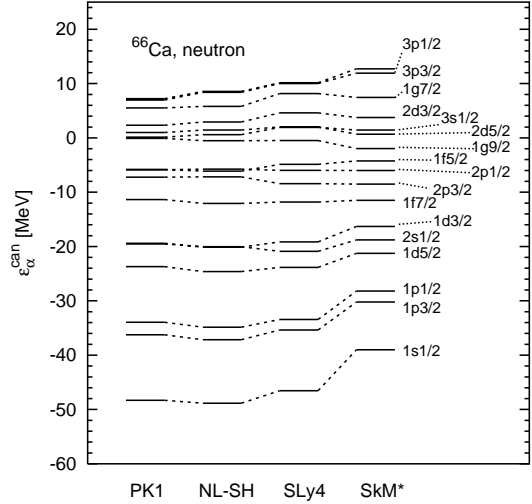}
\caption{\label{fig:levels}
\baselineskip=13pt
The neutron single particle level
$\varepsilon^{\rm can}_\alpha$ for $^{66}$Ca in RCHB calculations
with NLSH and PK1 and the Skyrme HFB calculations with SkM$^\ast$
and SLy4. }
\end{figure}

\section{\label{sec:66ca} single-particle structure}

To understand the halo phenomena and the different predictions in
RCHB and Skyrme HFB calculations, the nucleus $^{66}$Ca is taken as
an example. Figure \ref{fig:levels} depicts the single particle
level $\varepsilon^{\rm can}_\alpha$ in RCHB calculations with NLSH
and PK1 and the Skyrme HFB calculations with SkM$^\ast$ and SLy4
for neutrons in $^{66}$Ca, i.e., the diagonal elements of the
mean-field Hamiltonian in the canonical basis.

The order of the levels is essentially the same for all the
calculation. Furthermore the spectra for RCHB with PK1 and NL-SH are
quite similar. However, the level $1s_{1/2}$ in SkM$^\ast$ is
apparently higher than those of the other parameter sets. From left
to right, the shell gap become less apparent. The shell gaps at
$N=2$, 8, 20, 28, and 40 are clearly shown in the RCHB spectra,
while the shell gaps of $N=28$ and 40 do not appear in that of the
SkM$^\ast$ calculation (see also Fig.~\ref{fig:s2n}).

\begin{figure}
\includegraphics[width=0.45\textwidth]{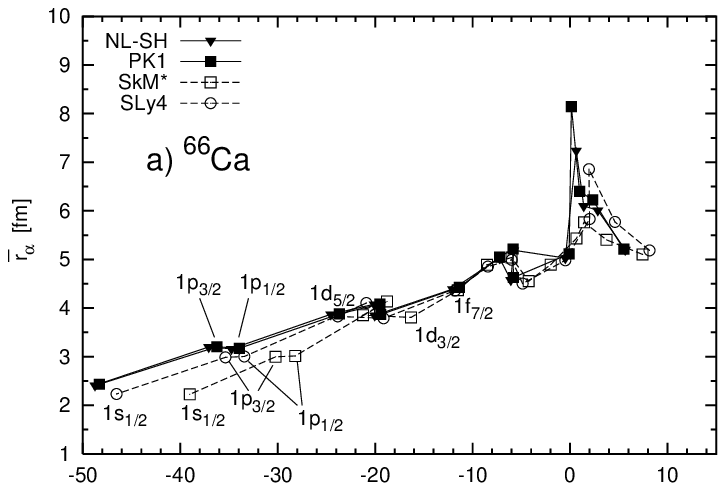}

\includegraphics[width=0.45\textwidth]{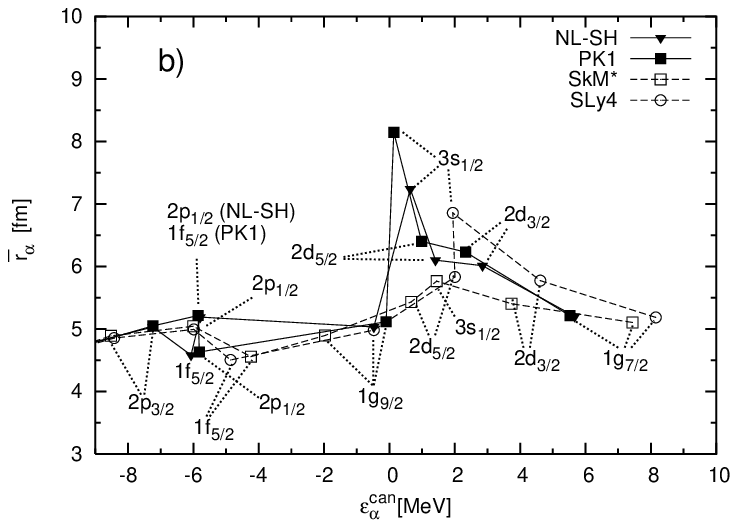}
\caption{\label{fig:r2_e}
\baselineskip=13pt
a) The $rms$ radii of each neutron
single-particle orbit in canonical basis $\bar{r}_\alpha$  as
functions of its single-particle energy $\varepsilon^{\rm
can}_\alpha$.  b) Same as a) but for the region around
$\varepsilon^{\rm can}=0$. }
\end{figure}

\begin{figure}
\includegraphics[width=0.33\textwidth]{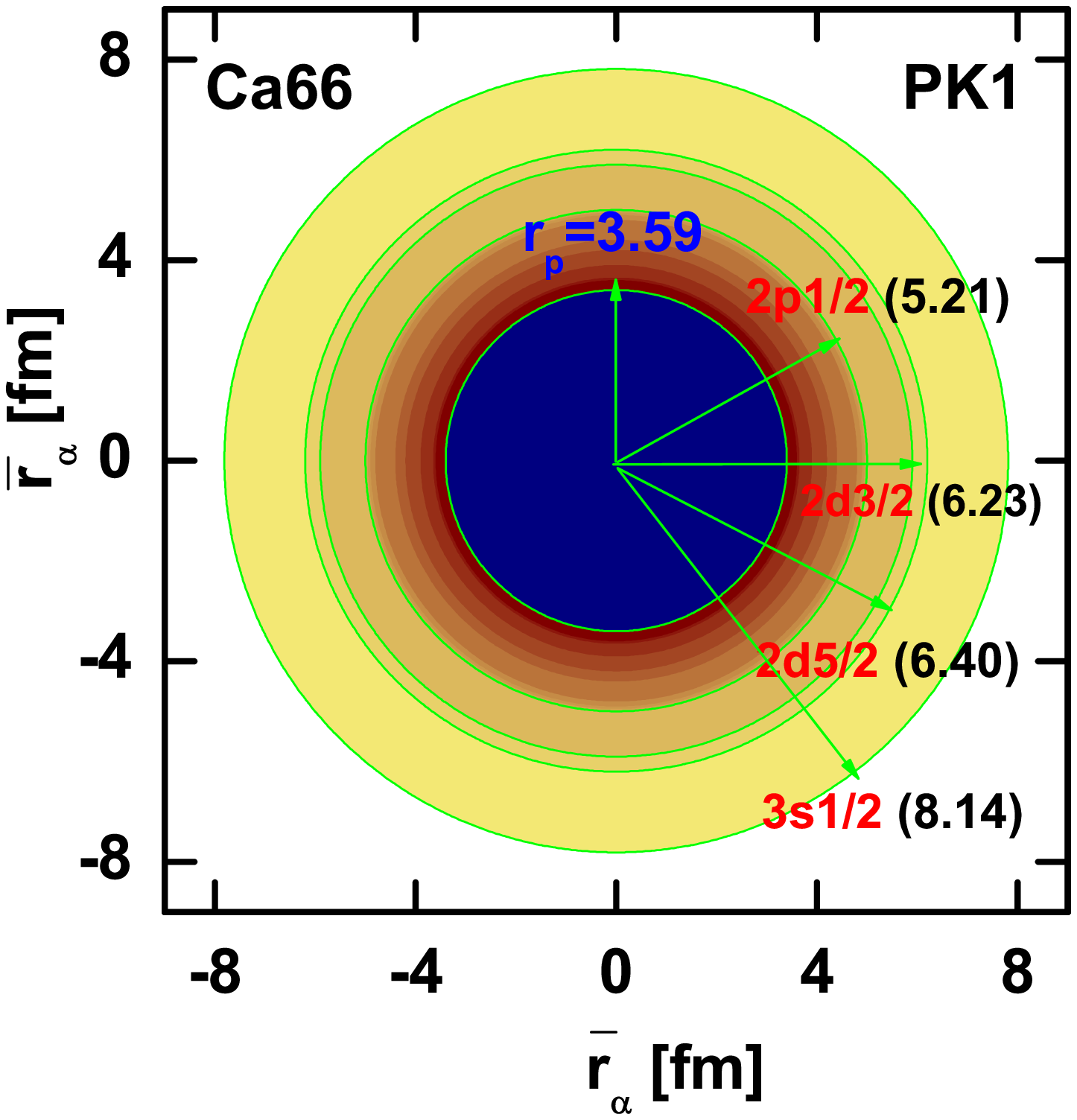}
\includegraphics[width=0.33\textwidth]{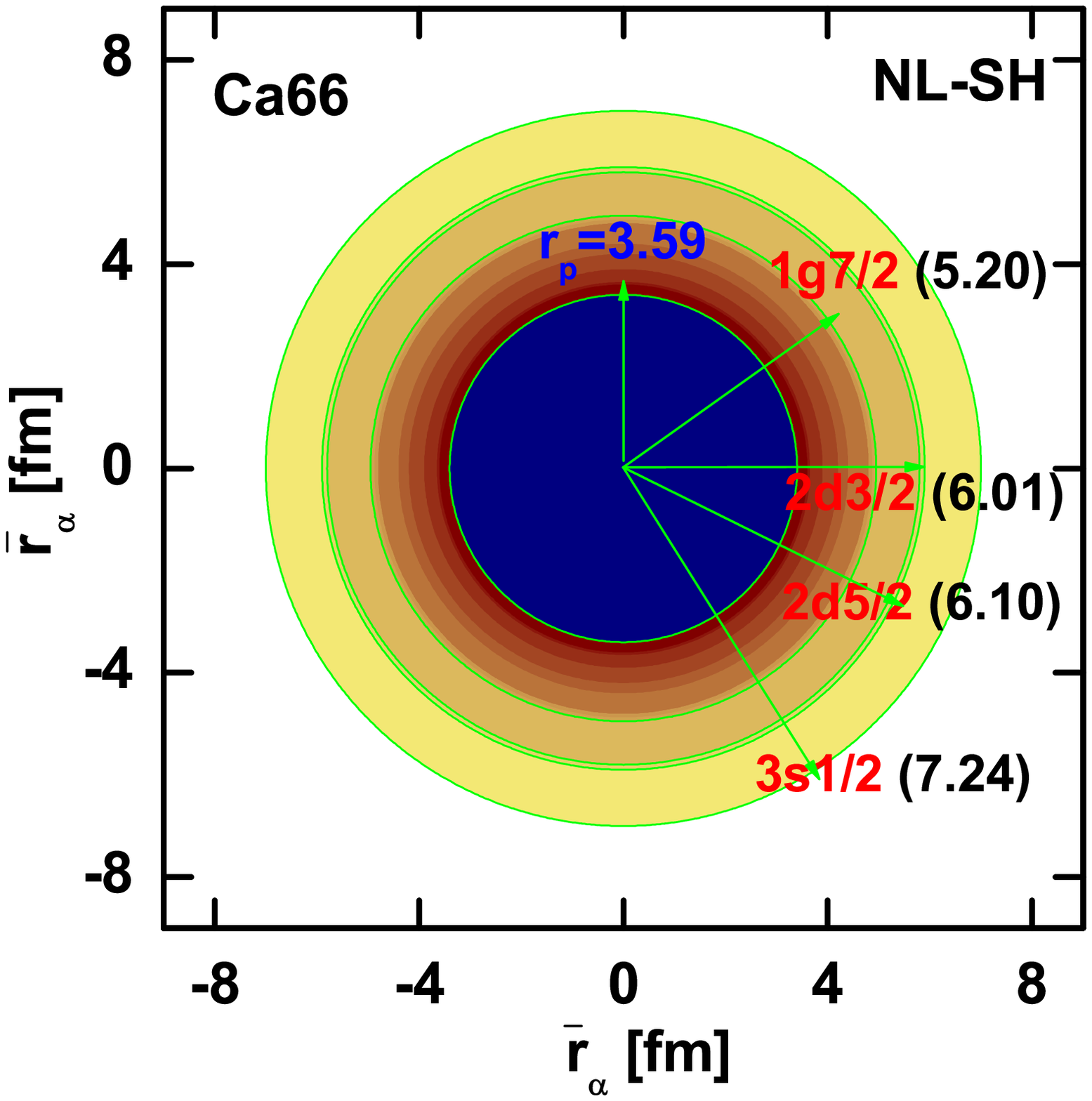}
\includegraphics[width=0.33\textwidth]{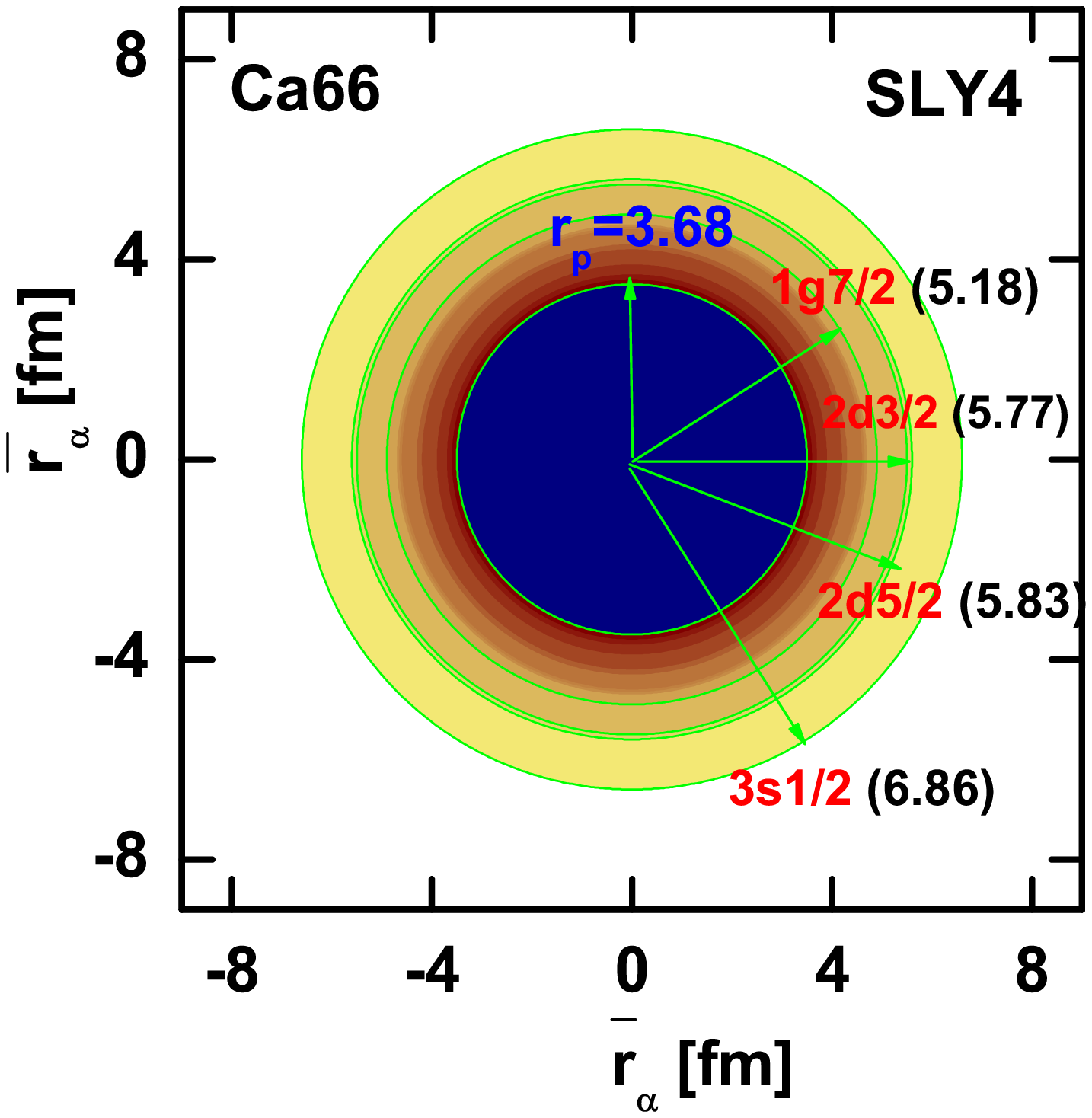}
\includegraphics[width=0.33\textwidth]{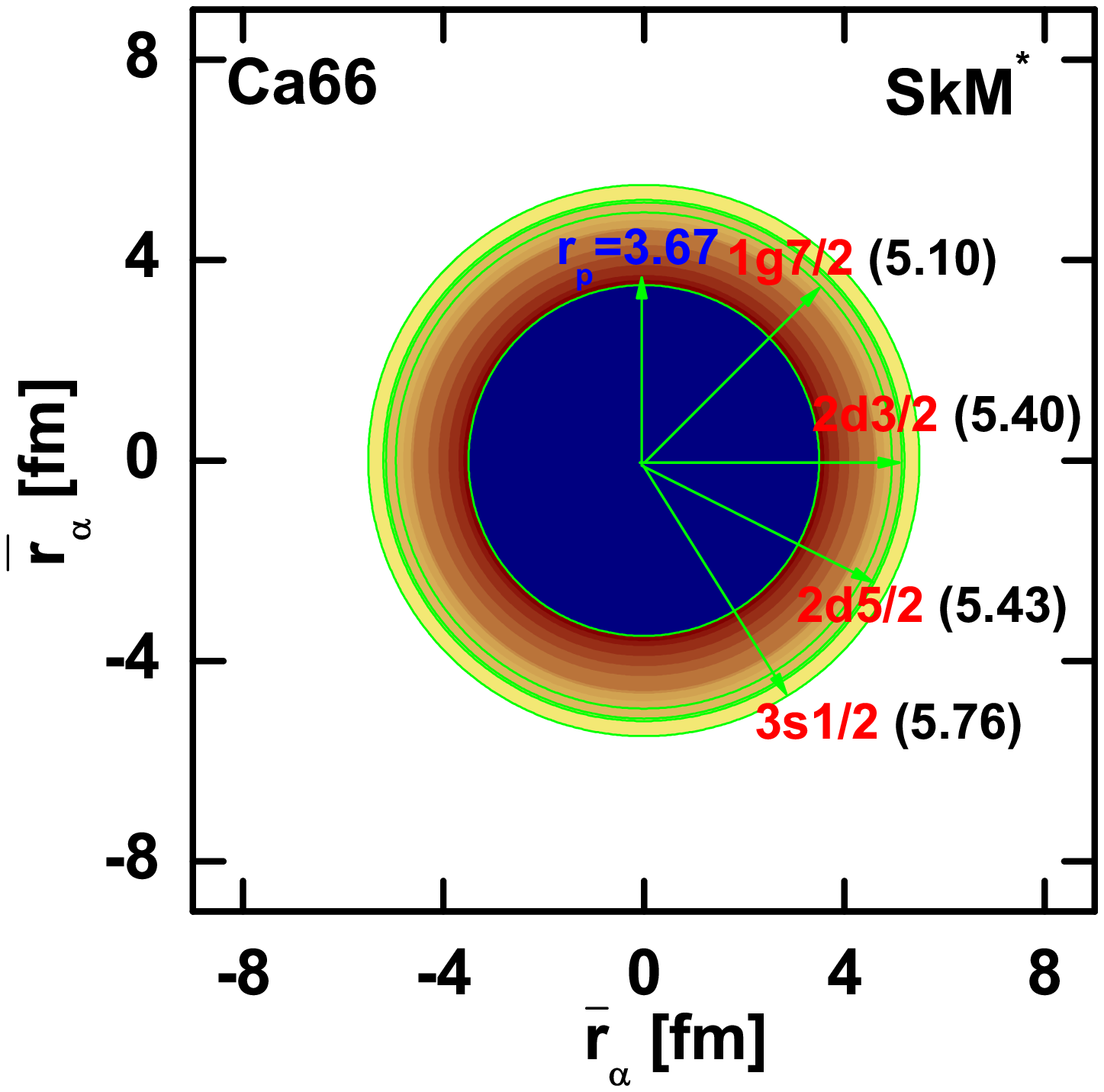}
\caption{\label{fig:r_circle}
\baselineskip=13pt
The  $rms$ radii of each neutron
single-particle orbit in canonical basis $\bar{r}_\alpha$ in RCHB
calculations with NLSH and PK1 and the Skyrme HFB calculations with
SkM$^\ast$ and SLy4 displayed as radii of circles. The proton one
$\bar{r}_{p}$ (the blue center circle in the on-line version) is
included for a comparison. The grade of color corresponds to the
order of $\bar{r}_\alpha$. }
\end{figure}

It has been pointed out \cite{Men02} that the neutron $3s_{1/2}$
level played an important role in the giant halo. Thus, comparison
is made in Fig.~\ref{fig:r2_e} for the $rms$ radii of each
single-particle orbit (canonical basis) $\bar{r}_\alpha$ in RCHB
calculations with NLSH and PK1 and the Skyrme HFB calculations with
SkM$^\ast$ and SLy4. The lines of $\bar{r}_\alpha$ versus
$\varepsilon^{\rm can}_\alpha$ are quite similar to each other
except for the $1s_{1/2}$ and the region around $\varepsilon^{\rm
can}=0$. This indicates the similarity of the potential wall except
for the bottom of the potential wall. A magnification around
$\varepsilon^{\rm can}=0$ is displayed in Fig.~\ref{fig:r2_e}b. The
most significant difference is $3s_{1/2}$: $\bar{r}_{3s1/2}$ of PK1
is more than 2 fm larger than that of SkM$^\ast$. These differences
are clearer in Fig.\ \ref{fig:r_circle} which illustrates
$\bar{r}_\alpha$ as radii of circles.

\begin{table}
\begin{center}
\hspace*{1em}
\vbox{\hsize=0.65\textwidth
\caption{\label{tab:radii}
\baselineskip=13pt
Properties ($\varepsilon^{\rm
can}_\alpha$, $\bar{r}_\alpha$, occupation probability $v_\alpha^2$,
and $\frac{2}{N}v_\alpha^2\bar{r}_\alpha$ ) of the neutron
$3s_{1/2}$ orbit and $\bar{r}_n$ of $^{66}$Ca. }
\begin{ruledtabular}
\begin{tabular}{lccccc}
\vspace*{-0.2em}
 & $\varepsilon^{\rm can}_{3s1/2}$ & $\bar{r}_{3s1/2}$ &
$v^2_{3s1/2}$ & $\frac{2}{N}v^2_{3s1/2}\bar{r}_{3s1/2}$ & $\bar{r}_n$ \\
 & [MeV] & [fm] & & [fm] & [fm] \\
\hline
\vspace*{-0.4em}
PK1          & 0.133 & 8.144 & 0.239 & 0.0846 & 4.363 \\
\vspace*{-0.4em}
NL-SH        & 0.640 & 7.240 & 0.089 & 0.0280 & 4.314 \\
\vspace*{-0.4em}
SLy4         & 1.939 & 6.856 & 0.016 & 0.0043 & 4.205 \\
\vspace*{-0.4em}
SkM$^{\ast}$ & 1.445 & 5.765 & 0.010 & 0.0025 & 4.210
\end{tabular}
\end{ruledtabular}
} 
\end{center}
\end{table}

Table \ref{tab:radii} shows $\varepsilon^{\rm can}_\alpha$,
$\bar{r}_\alpha$, occupation probability $v_\alpha^2$, and
 $\frac{2}{N}v_\alpha^2\bar{r}_\alpha$ for the neutron $3s_{1/2}$ orbit,
as well as $\bar{r}_n$ of $^{66}$Ca. Both $\bar{r}_{3s1/2}$ and
$v^2_{3s1/2}$ make $\bar{r}_n$ of PK1 the largest one. Since the
contribution of the orbit to $\bar{r}_n$ is given by
$\frac{2}{N}v^2_{3s1/2}\bar{r}_{3s1/2}$, it is seen from
Tab.~\ref{tab:radii} that 54 \% (0.082 fm) of the difference
$\bar{r}_n({\rm PK1})-\bar{r}_n({\rm SkM}^\ast)=0.153$ fm comes from
the neutron $3s_{1/2}$. It is also noted that the higher energy does
not necessarily mean the larger spatial distribution.

\begin{figure}
\includegraphics[width=0.45\textwidth]{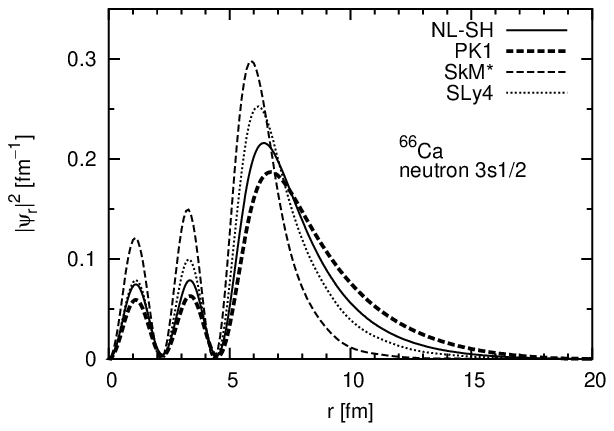}
\caption{\label{fig:wf2_3s1}
\baselineskip=13pt
Radial wave function squared $|\psi_r|^2$
(canonical basis) of the neutron $3s_{1/2}$ of $^{66}$Ca. They are
normalized as $\int dr\,|\psi_r|^2=1$.}
\end{figure}
\begin{figure}
\includegraphics[width=0.45\textwidth]{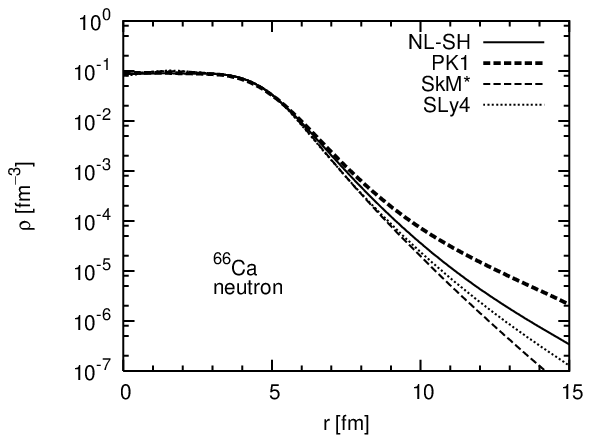}
\caption{\label{fig:denslog_n}
\baselineskip=13pt
Neutron density distribution
of $^{66}$Ca calculated with the four parameter sets. }
\end{figure}

The difference in the neutron $3s_{1/2}$ orbit
is shown also in Fig.~\ref{fig:wf2_3s1} which illustrates
the radial wave functions squared.
The tail of PK1 has appreciably longer distribution than the others, and
as a counter part, the amplitude of PK1 is smaller in the inner region.
The number and locations of the nodes are the same for the four lines.
The neutron density distributions have appreciable
difference  (Fig.~\ref{fig:denslog_n})
in accordance with the behavior of Fig.~\ref{fig:wf2_3s1}.

\begin{figure}
\includegraphics[width=0.45\textwidth]{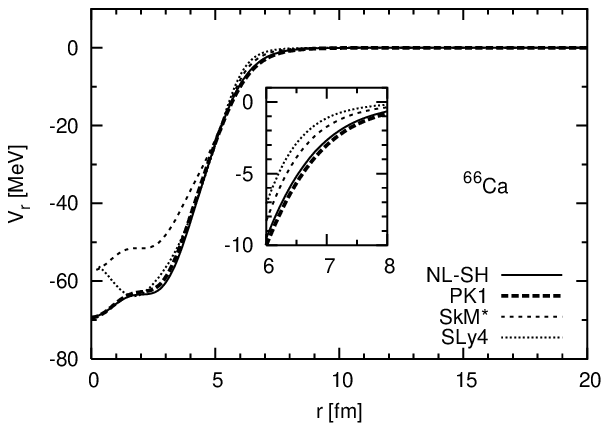}
\caption{\label{fig:pot}
\baselineskip=13pt
Central potential of the neutrons of $^{66}$Ca.
Inserted is a magnification of the surface region.}
\end{figure}
\begin{figure}
\includegraphics[width=0.45\textwidth]{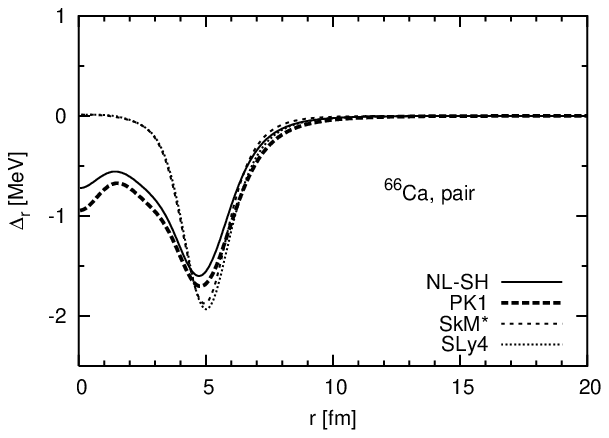}
\caption{\label{fig:pot_pair}
\baselineskip=13pt
Pairing  potential of the neutrons of $^{66}$Ca.}
\end{figure}

Figure \ref{fig:pot} depicts central potentials.
These are the summation of the scalar and the vector potentials
in the RCHB calculation, and the summation of the $t_0$ and $t_3$ terms
in the particle-hole potential in the Skyrme HFB calculation,
see Eq.~(A.5a) of Ref.~\cite{Dob84}.
A significant difference is that
the SkM$^\ast$ calculation has
smaller depth than the other calculations,
as is anticipated from Fig.~\ref{fig:levels}, and
another is a slight difference in the tail region (see the inset of
Fig.~\ref{fig:pot}).
It should be noted that
the tails of the RCHB calculations have longer distributions
than the Skyrme HFB calculations, and PK1 has the longest one.
Figure \ref{fig:pot_pair} displays the pairing potential
$$
\Delta_r = \frac{1}{2}V_0\left( 1-\frac{\rho(r)}{\rho_0}\right)
\tilde{\rho}_n(r),
$$
where $\rho(r)$ is the total nuclear density, and
$\tilde{\rho}_n(r)$ is the pairing density \cite{Dob84}
of neutron.
For the RCHB calculations, the $S=1$ component of the
pairing interaction is removed explicitly.
Again, the line of PK1 has the longest tail, therefore,
the large $\bar{r}_{3s1/2}$ of PK1 can be understood
in terms of the tails of the central and the pairing
potentials.

\begin{figure}
\includegraphics[width=0.45\textwidth]{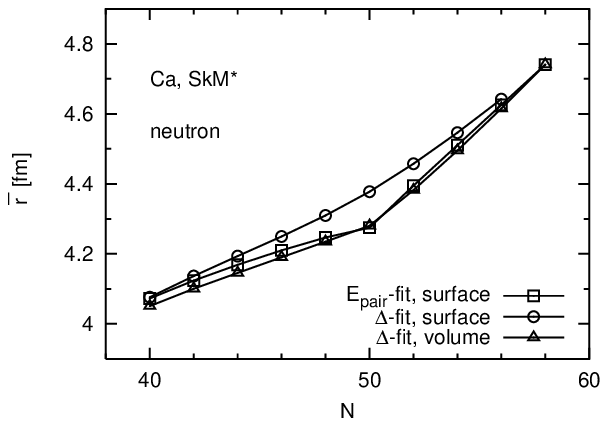}
\caption{\label{fig:r_test}
\baselineskip=13pt
$\bar{r}_n$ of the SkM$^\ast$ calculations with three pairing interactions:
the surface-type pairing with the strength determined from the
pairing energy of the RCHB calculations (squares),
the surface-type with the strength determined from the pairing gaps
(circles),
and the volume-type determined from the pairing gaps (triangles).
The lines of the second and the third ones are very close to each other.
}
\end{figure}

Finally, we mention the importance of the pairing correlations.
It turned out that when
the strength of the surface-type pairing interaction was determined
so as to reproduce the neutron pairing gaps
of $^{42,44,46,50}$Ca, 1.0--1.9 MeV, obtained from
the odd-even mass differences of the experimental data \cite{Aud03},
the pairing energy of the SkM$^\ast$ calculation was $-$(35--39) MeV
in $40 \leq N \leq 50$,
and  the kink of $\bar{r}_n$ at $N=50$ was completely smeared out
(the circles in Fig.~\ref{fig:r_test}).
It is noted that the pairing energy is more than 3 times larger than the
values of the RCHB calculations 8--9 MeV.
On the other hand, the volume-type pairing interaction
$V_0\delta({\bm r}-{\bm r^\prime})$
can reproduce the pairing energy of the RCHB calculation and
the experimental pairing gaps simultaneously, and the $\bar{r}_n$ has
the kink (the triangles in Fig.~\ref{fig:r_test}).
Therefore, the kink is strongly influenced by
the pairing interaction.
(See also Refs.~\cite{Miz00,Ben00}.)

\section{\label{sec:conclusion} Conclusion}
In this paper, the phenomena of the giant halo and halo of the neutron-rich
even-Ca isotopes have been investigated and compared in the framework of
the RCHB and the Skyrme HFB calculations.
With the two parameter sets for each of the RCHB and the Skyrme HFB
calculations, it has been found that although the halo phenomena existed for Ca
isotopes near the neutron drip line in both calculations,  the halo of
the Skyrme HFB calculations started at a more neutron-rich nucleus
than that of the RCHB calculations, and the RCHB calculations had
larger neutron rms radii systematically in $N
\geq 40$ than those of the Skyrme HFB calculations. The former
difference comes from difference in the shell structure. The reasons for
the latter can be partly explained by the neutron 3$s_{1/2}$ orbit,
which causes 50 \% of the difference
in the neutron rms radii among the four calculations at $^{66}$Ca.

\begin{acknowledgments}
This work is  partly supported by the National Natural Science
Foundation of China under Grant Nos.
10221003,
10435010,
10475003, and
10575036,
the Doctoral Program Foundation from the Ministry of Education in China,
 and the Knowledge Innovation
Project of Chinese Academy of Sciences under contract
Nos. KJCX2-SW-N02 and KJCX2-SW-N17.

\end{acknowledgments}

\vfill
\bibliography{c6}

\end{document}